\documentclass[aps,nofootinbib,prd,preprintnumbers]{revtex4}
\usepackage{amstext}
\usepackage{amssymb}
\usepackage{amsmath}
\usepackage{graphicx}
\usepackage{rotating}
\usepackage{gensymb}
\usepackage[format=plain,singlelinecheck = false, format= hang, justification=raggedright]{caption}

\usepackage{multirow} 
\usepackage[T1]{fontenc} 
\usepackage{slashed}
\usepackage{caption}
\usepackage{float}
\usepackage[normalem]{ulem}

\usepackage[usenames,dvipsnames]{color}
\definecolor{mightnightblue}{RGB}{25,25,112}
\usepackage{hyperref}
 \hypersetup{
     colorlinks=true,
     linkcolor= Black,
     citecolor=mightnightblue,
     urlcolor=mightnightblue
     } 

\def\gsim{\raise0.3ex\hbox{$\;>$\kern-0.75em\raise-1.1ex\hbox{$\sim\;$}}}
\def\lsim{\raise0.3ex\hbox{$\;<$\kern-0.75em\raise-1.1ex\hbox{$\sim\;$}}}
\newcommand {\ignore}[1]{}

\newcommand{\eVq}  {\text{eV}^2}

\newcommand{\AddrAHEP}{%
  AHEP Group, Institut de F\'{i}sica Corpuscular --
  C.S.I.C./Universitat de Val\`{e}ncia, Parc Cientific de Paterna.\\
 C/ Catedratico Jos\'e Beltr\'an, 2 E-46980 Paterna (Val\`{e}ncia) - SPAIN}

\begin{document}

\begin{flushright}
FTUV-17-11-29, IFIC/17-57 \\
\end{flushright}

\title{Neutrinos, DUNE and the world best bound on CPT invariance}

\author{G. Barenboim$^1$}\email{Gabriela.Barenboim@uv.es}
\author{C. A. Ternes$^2$}\email{chternes@ific.uv.es}
\author{M. T{\'o}rtola~$^2$}\email{mariam@ific.uv.es}  
\affiliation{$^1$~Departament de F{\'i}sica Te{\'o}rica and IFIC, Universitat de Val{\`e}ncia-CSIC, E-46100, Burjassot, Spain}
\affiliation{$^2$~\AddrAHEP}


\vskip 2cm

\begin{abstract}

CPT symmetry, the combination of Charge Conjugation, Parity and Time reversal, is a cornerstone of our model building 
strategy and therefore the repercussions of its potential violation will severely threaten the most extended tool 
we currently use to describe physics, {\it i.e.} local relativistic quantum fields. However, limits on its conservation from 
the Kaon system look indeed imposing. In this work we will show that neutrino oscillation experiments can improve this limit by 
several orders of magnitude and therefore are an ideal tool to explore the foundations of our approach to Nature. 

Strictly speaking testing  CPT violation would require an explicit model for how CPT is broken and its effects on physics. Instead, what is presented in this paper is a  test of one of the predictions of CPT conservation, i.e., the same mass and mixing parameters in neutrinos and antineutrinos. 
In order to do that we calculate the current CPT bound on all the neutrino mixing parameters and study the sensitivity of the DUNE experiment to such an observable. 
After deriving the most updated bound on CPT from neutrino oscillation data,  we show that, if the recent T2K results turn out to be the true values of neutrino and antineutrino oscillations, DUNE would measure the fallout of
CPT conservation at more than 3$\sigma$. Then, we study the sensitivity of the experiment to measure CPT invariance in 
general, finding that DUNE will be able to improve the current bounds on $\Delta(\Delta m^2_{31})$ by at least one order of magnitude. 
We also study the sensitivity to the other oscillation parameters. Finally we show that, if CPT is violated in nature,  
combining neutrino with antineutrino data in oscillation analysis will produce imposter solutions.

\end{abstract}
\maketitle

\section{Introduction}
\label{sec:intro}
CPT invariance is arguably one of the few sacred cows of particle physics. Its position as such arises from the fact that
CPT conservation is a natural consequence of only three assumptions: Lorentz invariance, locality and hermiticity of the Hamiltonian, 
all of which have plenty of reasons to be included in our theory, besides CPT itself. In short, the CPT theorem states that particle and 
antiparticle have the same mass and, if unstable, also the same lifetime (for a nice proof of the CPT theorem see Ref.~\cite{Streater:1989vi}).
Therefore, the consequences of finding evidence of CPT non-conservation would be gigantic \cite{Barenboim:2002tz}. At least one of the three ingredients above must be false and our model building strategy would need to be revisited.

It should be noted however that testing the predictions of CPT conservation is not strictly equivalent to constraining CPT violation. 
Tests of CPT conservation might be performed by comparing the masses of particles and antiparticles. Indeed, these mass differences might be regarded as CPT violating observables. Nevertheless, the interpretation and comparison of bounds from different observables would only be possible with the consideration of a particular model of CPT violation.

Having said that, it is also clear that tests of CPT invariance have been historically   associated with the neutral kaon system and therefore although in the absence of an explicit model any connection is meaningless, the comparison between kaons and neutrinos seems unavoidable.
A superficial face value extrapolation leaves no room to be optimistic: the current limits on CPT violation arising from the neutral Kaon system seem to be quite solid 
\begin{equation}
  \frac{|m(K^0) - m(\overline{K}^0)|}{m_K} < 0.6 \times 10^{-18}\,. 
  \label{eq:mK}
\end{equation}

However, the strength of this limit is indeed artificial. Its robustness derives from the choice of the scale in the denominator, which 
is arbitrary at any rate and has nothing to do with a concrete model of CPT violation. Besides 
the Kaon is not an elementary particle and therefore this test has more to do with testing QCD rather than a fundamental symmetry 
of (elementary) fermions. Additionally, the parameter present in the Lagrangian is not the mass but the mass squared 
and therefore this limit should be re-written as
\begin{equation}
  |m^2(K^0) - m^2(\overline{K}^0)| < 0.25~\mbox{eV}^2 \,.
  \label{eq:mK2}
\end{equation}
Now it becomes obvious that neutrino experiments can test CPT to an unprecedented extent and therefore
can provide stronger limits than the ones regarded as the most stringent now\footnote{CPT was tested also using charged leptons. However, these measurements involve a combination
of mass and charge and are not a direct CPT test. Only neutrinos can provide CPT tests on an elementary mass not contaminated by charge.}. 
Let us stress again, however, that without an explicit model for CPT violation it is not straightforward or even meaningful  to compare the neutrino-antineutrino mass squared differences and the kaon ones. CPT violation may show up only in one of the sectors and therefore the strong bounds in one of them might not be directly applicable to the other. 

On the other hand, there are reasons to believe neutrinos are an ideal
probe for CPT violation: quantum gravity is assumed to be non-local, opening the door to a potential CPT violation. Its effects however are expected to be 
Planck suppressed, {\it i.e.} $\left\langle v\right\rangle^2/M_{\text{P}} $, exactly in the right ballpark for neutrino experiments to see them.

\begin{figure}[t!]
 \centering
        \includegraphics[width=\textwidth]{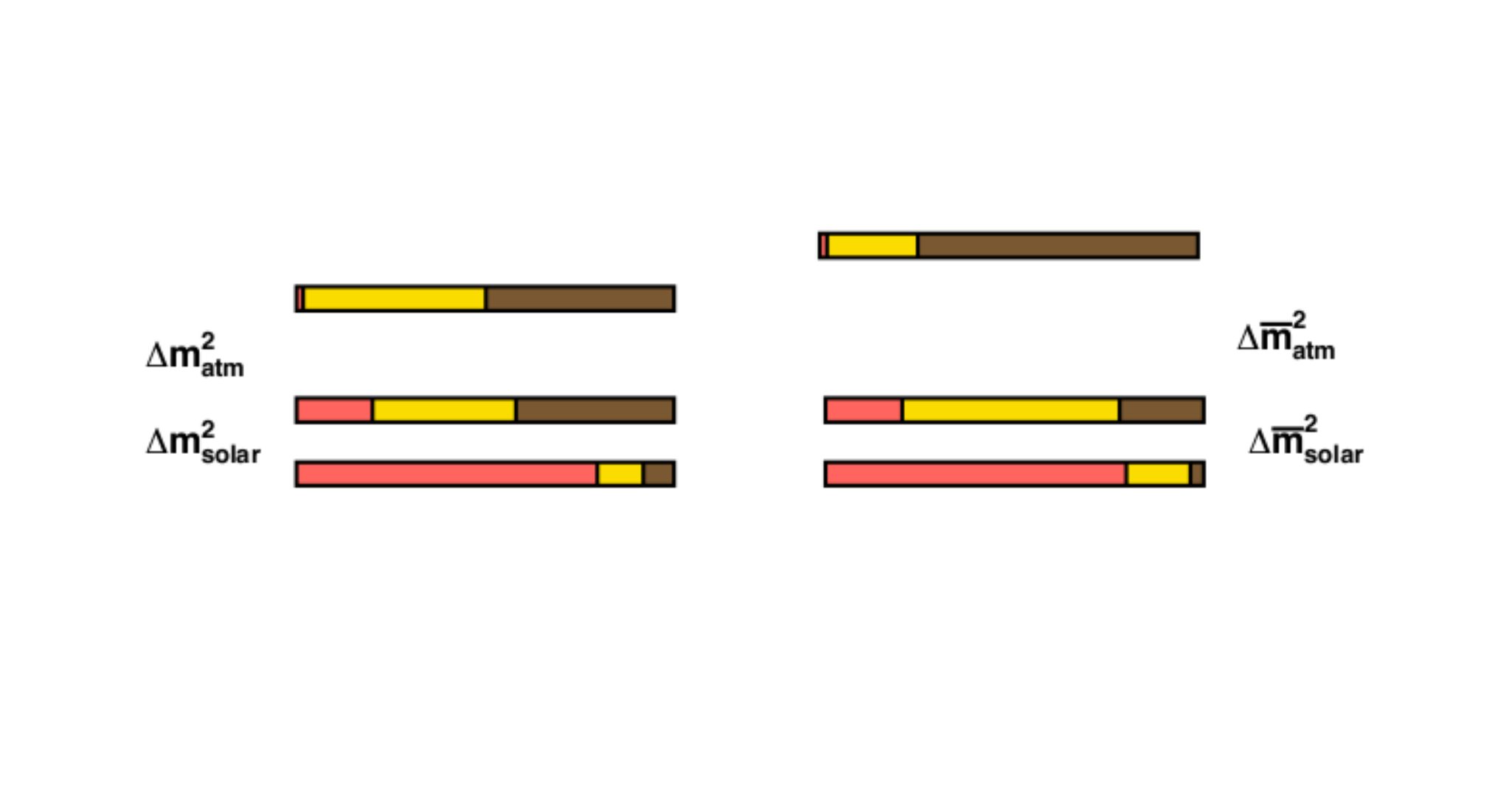}
       \captionsetup{justification=raggedright}
        \caption{Generic CPT violating spectrum. We have not included an overall shift between the neutrino and antineutrino sector as 
        it cannot be tested by oscillation experiments.}
	\label{fig:CPT-spectrum}
\end{figure}

Furthermore, as it is well known, neutrinos offer a unique mass generation mechanism, the see-saw, and therefore their masses
are sensitive to new physics and new scales. Scales where non-locality can be expected to show up. Of course,  in lack of a 
concrete theory of flavor, let alone one of CPT violation, the difference in the spectra of neutrinos and antineutrinos can appear
not only in the mass eigenstates but also in the mixing between flavor and mass eigenstates. Neutrino oscillation experiments can test only
CPT in the mass differences and mixing angles. An overall shift on the spectrum of neutrinos 
relative to that of antineutrinos
cannot be detected in oscillation experiments and can be bound only by cosmological data, see Ref.~\cite{Barenboim:2017vlc}.
It is important to notice that future kinematical direct searches for neutrino mass use only antineutrinos and thus 
cannot be used as a CPT test on the absolute mass scale either. (See Fig.~\ref{fig:CPT-spectrum}.)

Studies separating neutrinos and antineutrinos were done in the past \cite{Barenboim:2009ts,Adamson:2013whj,Abe:2017bay,Abe:2011ph} 
under several assumptions. In Ref.~\cite{Ohlsson:2014cha} the authors obtained the 
following model-independent bounds on CPT invariance  for the different parameters~\footnote{Here we follow the standard convention of denoting neutrino parameters as $\Delta m_{ij}^2$, $\theta_{ij}$, and antineutrino parameters as $\Delta \overline{m}_{ij}^2$, $\overline{\theta}_{ij}$.}:
 \begin{eqnarray}
  \label{eq:Ohlsson-bounds}
 & |\Delta m_{21}^2-\Delta \overline{m}_{21}^2| &< 5.9\times 10^{-5} \, \text{eV}^2,
  \nonumber \\
  & |\Delta m_{31}^2-\Delta \overline{m}_{31}^2| &< 1.1\times 10^{-3} \, \text{eV}^2,
 \nonumber \\
  & |\sin^2\theta_{12}-\sin^2\overline{\theta}_{12}| &< 0.25,
  \\
  & |\sin^2\theta_{13}-\sin^2\overline{\theta}_{13}| &< 0.03,
  \nonumber \\
  & |\sin^2\theta_{23}-\sin^2\overline{\theta}_{23}| &< 0.44 \nonumber ,
 \end{eqnarray}
at 3$\sigma$. MINOS has also bounded the difference in the atmospheric mass-splitting to be
\begin{equation}
 |\Delta m_{31}^2-\Delta \overline{m}_{31}^2| < 0.8\times 10^{-3} \, \text{eV}^2 
 \label{eq:MINOS}
\end{equation}
at 3$\sigma$, see Ref.~\cite{Adamson:2013whj}. Although this latter bound is stronger than the one in Eq.~(\ref{eq:Ohlsson-bounds}), 
it is not indicated whether it has been obtained after marginalizing over the atmospheric mixing angle or not. 
In any case, it seems clear that  the previous bounds in Eqs.~(\ref{eq:Ohlsson-bounds}) and (\ref{eq:MINOS})  have been 
derived assuming the same mass ordering for neutrinos and antineutrinos. Note that different mass orderings for neutrinos and 
antineutrinos would automatically imply CPT violation, even if the same value for the mass difference is obtained.
At this point it is worth noting  that, in this work, we are not considering any particular model of CPT violation and therefore all the results obtained can be regarded as model-independent. 

In the light of the new experimental data, mainly from reactor and long--baseline accelerator experiments, here we are going to update the bounds 
on CPT from neutrino oscillation data. We will use basically the same data considered in the global fit to neutrino oscillations in  
Ref.~\cite{deSalas:2017kay}.   Note, however, that in this work we will analyze neutrino and antineutrino data separately. 
Given that current atmospheric experiments, such as 
Super-Kamiokande~\cite{Abe:2017aap}, IceCube-DeepCore~\cite{Aartsen:2014yll,Aartsen:2017nmd} and ANTARES~\cite{AdrianMartinez:2012ph},
can not  distinguish neutrinos from antineutrinos event by event,  we will not include them in this study. 
Here we summarize the neutrino samples considered, indicating in each case the neutrino or antineutrino parameters they are sensitive to
\begin{itemize}
 \item  solar neutrino data~\cite{Cleveland:1998nv,Kaether:2010ag,Abdurashitov:2009tn,hosaka:2005um,Cravens:2008aa,Abe:2010hy,Nakano:PhD,Aharmim:2008kc,Aharmim:2009gd,Bellini:2013lnn}:  $\theta_{12}$, $\Delta m_{21}^2$, $\theta_{13}$
 \item neutrino mode in long--baseline experiments K2K~\cite{Ahn:2006zza}, MINOS~\cite{Adamson:2013whj,Adamson:2014vgd}, T2K~\cite{Abe:2017uxa,Abe:2017bay} and NO$\nu$A~\cite{Adamson:2017qqn,Adamson:2017gxd}:  $\theta_{23}$, $\Delta m_{31}^2$, $\theta_{13}$
 \item  KamLAND reactor antineutrino data~\cite{Gando:2010aa}: $\overline{\theta}_{12}$, $\Delta \overline{m}_{21}^2$, $\overline{\theta}_{13}$
 \item short--baseline reactor antineutrino experiments Daya Bay~\cite{An:2016ses}, RENO~\cite{RENO:2015ksa} and Double Chooz~\cite{Abe:2014bwa}:     $\overline{\theta}_{13}$, $\Delta \overline{m}_{31}^2$
\item antineutrino mode in long--baseline experiments\footnote{The K2K experiment took only data in neutrino mode. The NO$\nu$A experiment has not yet published data in antineutrino mode.}    MINOS~\cite{Adamson:2013whj,Adamson:2014vgd} and T2K~\cite{Abe:2017uxa,Abe:2017bay}: $\overline{\theta}_{23}$, $\Delta \overline{m}_{31}^2$
$\overline{\theta}_{13}$           
\end{itemize}

There is no reason to put bounds on $|\delta-\overline{\delta}|$ at the moment, since all possible values of
$\delta$ or $\overline{\delta}$ are allowed. The exclusion of certain values of $\delta$ in Ref.~\cite{deSalas:2017kay} can only be obtained 
after combining neutrino and antineutrino data. Hence, performing such an exercise, the most up-to-date bounds on CPT violation are:
 \begin{eqnarray}
 & |\Delta m_{21}^2-\Delta \overline{m}_{21}^2| &< 4.7\times 10^{-5} \text{eV}^2,
  \nonumber \\
  & |\Delta m_{31}^2-\Delta \overline{m}_{31}^2| &< 3.7\times 10^{-4} \text{eV}^2,
 \nonumber \\
  & |\sin^2\theta_{12}-\sin^2\overline{\theta}_{12}| &< 0.14,
  \\
  & |\sin^2\theta_{13}-\sin^2\overline{\theta}_{13}| &< 0.03,
  \nonumber \\
  & |\sin^2\theta_{23}-\sin^2\overline{\theta}_{23}| &< 0.32\nonumber ,
 \label{eq:new-bounds}
 \end{eqnarray} 
improving the older bounds in Eqs.~(\ref{eq:Ohlsson-bounds}) and (\ref{eq:MINOS}), except for $\sin^2\theta_{13}$, that remains unchanged.
 Note that the limit on $\Delta(\Delta m_{31}^2)$  is already better than the one of the neutral Kaon system and 
should be regarded as the best bound on CPT violation on the mass squared so far. It should be noted as well that, to obtain these bounds we assume that neutrinos and 
antineutrinos have the same definition of $\Delta m^2 $, {\it i.e.} the mass difference has the same sign. In principle, of course the mass difference in neutrinos and antineutrinos may have a different sign, but in this case one may argue that the sign difference is already a sign of CPT violation in itself.

Our article is structured as follows: in section~\ref{sec:simulation} we explain the details of our simulation of the DUNE experiment. 
In section~\ref{sec:t2k}, we consider the independent analysis of neutrino and antineutrino data performed by the T2K collaboration 
and analyze the sensitivity of DUNE to this scenario with neutrino and antineutrino parameters fixed by the T2K analysis. 
In section~\ref{sec:sensitivity} we check DUNE's sensitivity to measure CPT violating effects in all  the oscillation parameters, 
except the solar ones, to which DUNE will have no sensitivity. 
In section~\ref{sec:impost} we explicitly show  that,  by  performing the joined analysis of neutrino and antineutrino data, fake 
solutions, which we dubbed imposter solutions, can be obtained evidencing that the separate analysis is not
an option. It is a must. Otherwise one risks sacrificing the physics for the sake of statistics.
 Finally in section~\ref{sec:summary} we summarize our results and give some concluding remarks.

\section{Simulation of the DUNE experiment}
\label{sec:simulation}
The Deep Underground Neutrino Experiment (DUNE) will consist of two detectors exposed to a megawatt-scale muon neutrino beam
that will be produced at Fermilab. One detector will be placed near the source of the beam, while a second, much larger, detector 
comprising four 10-kiloton liquid argon TPCs will be installed 
1300 kilometers away of the neutrino source.  The primary scientific goal of DUNE is
the precision measurement of the parameters that govern neutrino mixing.
To simulate DUNE we use the GLoBES package \cite{Huber:2004ka,Huber:2007ji} with the most recent DUNE configuration file provided by 
the Collaboration \cite{Alion:2016uaj} used to produce the plots in Ref.~\cite{Acciarri:2015uup}. We assume DUNE to be running 3.5 years in neutrino mode and another 3.5 years 
in antineutrino mode. Assuming an 80 GeV beam with 1.07 MW beam power, this corresponds to 
an exposure of 300 kton--MW--years. 
This means that, in this configuration, DUNE will be using $1.47\times 10^{21}$ protons on target (POT) per year, which amounts basically in one single year to the same amount T2K has used in all of its lifetime until now (runs 1--7c)~\cite{Abe:2017uxa}. Our analysis includes disappearance and appearance 
channels, simulating signals and backgrounds. The simulated backgrounds include contamination of antineutrinos (neutrinos) in the neutrino (antineutrino) mode,  and also misinterpretation of flavors. Unfortunately, using GLoBES has one  disadvantage in the treatment of backgrounds,   since for a given channel, for instance the neutrino channel, the antineutrino backgrounds are oscillated with the same  probability as the neutrino signals and vice versa. While it is possible to use a customized probability engine in GLoBES, we have actually checked that the effect of the backgrounds is negligible. Therefore, in our analysis  we oscillate the backgrounds with the same parameters as the signal. 
In any case, in order to mitigate the impact such a simplification can have, we have increased the systematic errors due to misidentification of neutrinos by antineutrinos and vice versa  
 by a further 25\% over the original error given by the collaboration. 
Note, however, that this limitation  would only potentially affect the study in section~\ref{sec:t2k}, since for the sensitivity studies we always assume all of the parameters to be equal
for neutrinos and antineutrinos. 

\begin{table}[H]\centering
   \begin{tabular}{lc}
    \hline
    parameter & value 
    \\
    \hline
    $\Delta m_{31}^2$ & 2.60$\times 10^{-3} \,\eVq$ \\
    $\Delta\overline{m}_{31}^2$  &2.62$\times 10^{-3} \,\eVq$ \\ 
    $\sin^2\theta_{23}$ & 0.51 \\
$\sin^2\overline{\theta}_{23}$ & 0.42\\[2mm]
\hline\\[-3mm]
    $\Delta m^2_{21}$, $\Delta\overline{m}^2_{21}$& $7.56\times 10^{-5} \,\eVq$\\  
    $\sin^2\theta_{12}$, $\sin^2\overline{\theta}_{12}$ & 0.321\\ 
     $\sin^2\theta_{13}$, $\sin^2\overline{\theta}_{13}$& 0.02155\\
   $\delta$, $\overline{\delta}$  & 1.50$\pi$\\
       \hline
     \end{tabular}
     \captionsetup{justification=centering}
     \caption{Oscillation parameters considered in the analysis of Sec.~\ref{sec:t2k}.}
     \label{tab:par1} 
\end{table}

\section{Probing the T2K neutrino and antineutrino analysis in DUNE}
\label{sec:t2k}

In this section we explore the sensitivity of DUNE to the separate analysis of neutrino and antineutrino data performed by the 
T2K Collaboration in Ref.~\cite{Abe:2017bay}. Therefore, we consider the best fit values in this analysis as the true values for
the atmospheric  parameters:
 $\Delta m_{31}^2=2.60\times 10^{-3}\, \eVq$ and $\sin^2\theta_{23}=0.51$ for neutrinos and 
 $\Delta \overline{m}_{31}^2=2.62\times 10^{-3} \,\eVq$ and 
$\sin^2\overline{\theta}_{23}=0.42$ for antineutrinos. The analysis considers only normal mass ordering, as we assume that the
current hint on this being the path followed by Nature will be solid when DUNE turns on.
The remaining oscillation parameters are fixed to their best fit value from the current global fit of neutrino oscillations 
in Ref.~\cite{deSalas:2017kay} except for the CP-violating phase, 
set to $\delta=\overline{\delta}=3\pi/2$ for simplicity. 
The neutrino and antineutrino oscillation parameters used in this study are summarized in Table~\ref{tab:par1}. Note that we assume the neutrino oscillations being parameterized by the usual PMNS matrix $U_{\text{PMNS}}$, with parameters $\theta_{12},\theta_{13},\theta_{23},\Delta m_{21}^2,\Delta m_{31}^2,\delta$, while the antineutrino oscillations are parameterized by a matrix $\overline{U}_{\text{PMNS}}$ with parameters $\overline{\theta}_{12},\overline{\theta}_{13},\overline{\theta}_{23},\Delta \overline{m}_{21}^2,\Delta \overline{m}_{31}^2,\overline{\delta}$. This results in the same probability functions for antineutrinos as for neutrinos with the neutrino parameters replaced by their antineutrino counterparts, besides the standard change of sign in the CP phase.

\begin{figure}[t!]
 \centering
        \includegraphics[width=\textwidth]{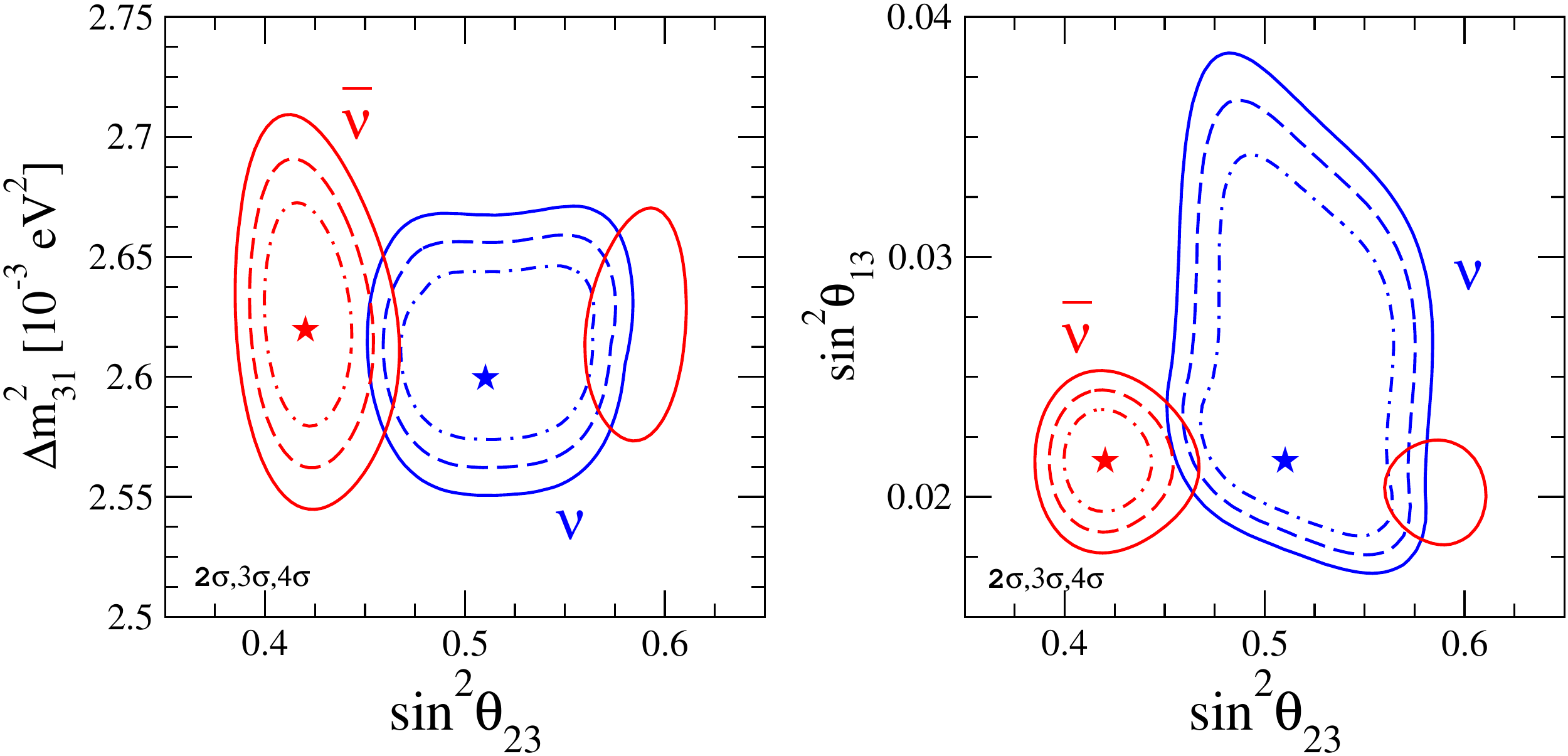}
       \captionsetup{justification=raggedright}
        \caption{DUNE  expected regions at 2$\sigma$, 3$\sigma$ and 4$\sigma$ in the atmospheric plane 
        $\sin^2\theta_{23}$ -- $\Delta m^2_{31}$ (left) and the $\sin^2\theta_{23}$ -- $\sin^2\theta_{13}$  plane (right) for 
        neutrinos (blue) and antineutrinos (red). The stars indicate the assumed true values for neutrino (blue) and antineutrino (red) oscillation parameters.}
	\label{fig:sq23-dm31-sq13}
\end{figure}

\begin{figure}[h!]
 \centering
        \includegraphics[width=0.9\textwidth]{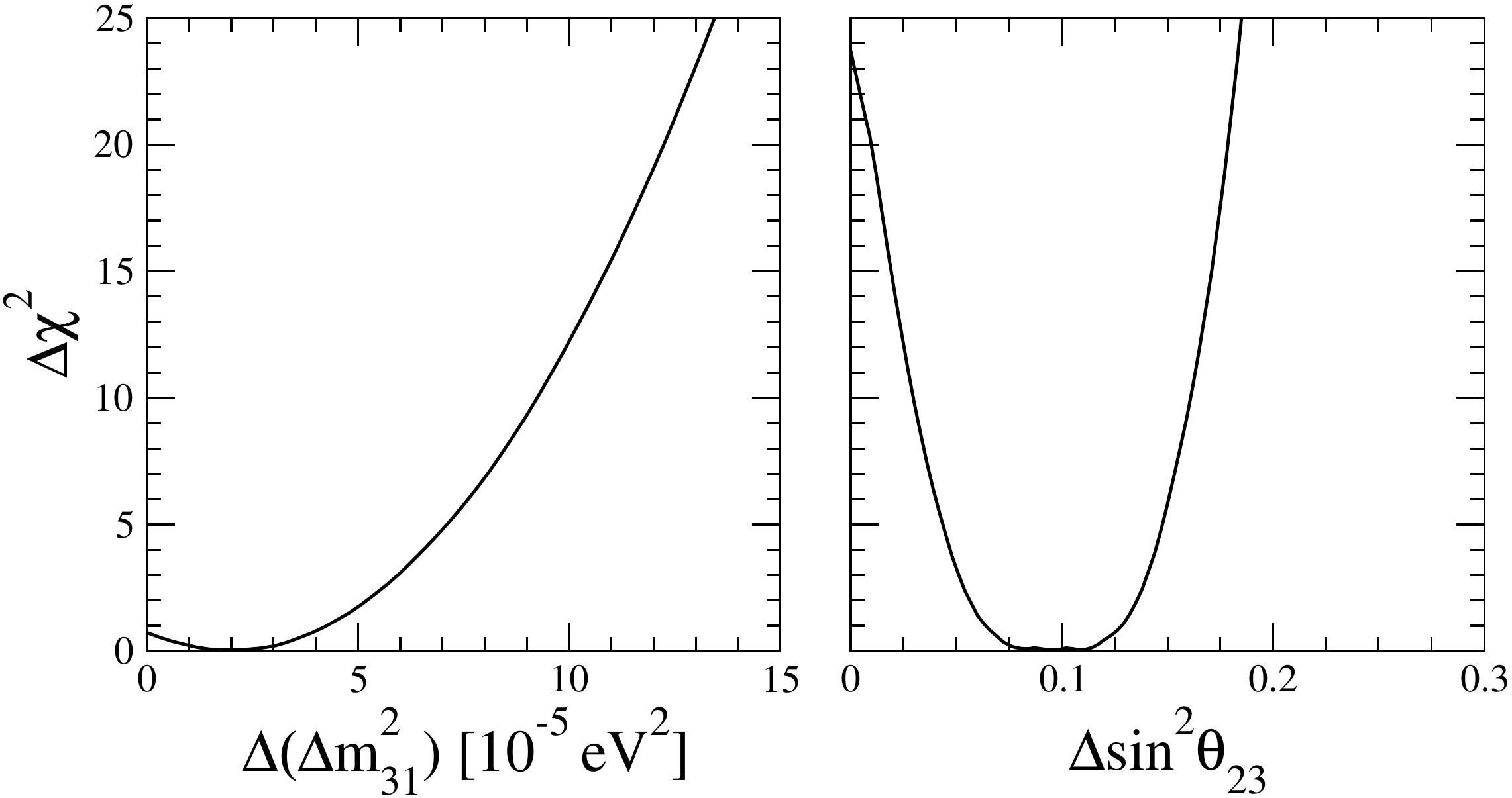}
       \captionsetup{justification=raggedright}
        \caption{ $\Delta\chi^2$ profiles as a function of $\Delta(\Delta m^2_{31})$ and $\Delta\sin^2\theta_{23}$ 
        assuming the parameters obtained by T2K as true values.}
	\label{fig:sens-CPT-t2k-sq23-dm31}
\end{figure}

We then simulate DUNE neutrino and antineutrino mode data using the parameters above as the true values and try to reconstruct them 
within the sensitivity of the DUNE experiment.
In the antineutrino channel, a prior on the determination of the reactor mixing angle, 
$\sin^2\overline{\theta}_{13}=0.02155\pm0.00090$~\cite{deSalas:2017kay}, is considered. 
This result comes mainly from the latest measurements of  the Daya Bay reactor experiment~\cite{An:2016ses}. 
We present the results of the analysis of neutrino and antineutrino data together in the same plot, projecting over two-dimensional regions and
 marginalizing over the other parameters not plotted. In Fig.~\ref{fig:sq23-dm31-sq13} (left) we present the allowed regions  
 at 2$\sigma$, 3$\sigma$ and 4$\sigma$ in the atmospheric plane ($\sin^2\theta_{23}$, $\Delta m^2_{31}$). There one can see that, 
 unlike what happens for T2K, in DUNE there would be no overlap of the allowed regions at the 3$\sigma$ 
level, although there would be still some overlap at the 4$\sigma$ level. 
To make this point even clearer,  we have plotted in Fig.~\ref{fig:sens-CPT-t2k-sq23-dm31} the sensitivity to $\Delta(\Delta m_{31}^2)=|\Delta m_{31}^2-\Delta \overline{m}_{31}^2|$ and $\Delta\sin^2\theta_{23}=|\sin^2\theta_{23}-\sin^2\overline{\theta}_{23}|$ from this analysis. There one can see that for the mass splittings  CPT conservation remains allowed at 1$\sigma$, while for the mixing angles the hypothesis of CPT conservation is disfavored at  around the 5$\sigma$ level.

\begin{figure}[ht!]
 \centering
        \includegraphics[width=0.9\textwidth]{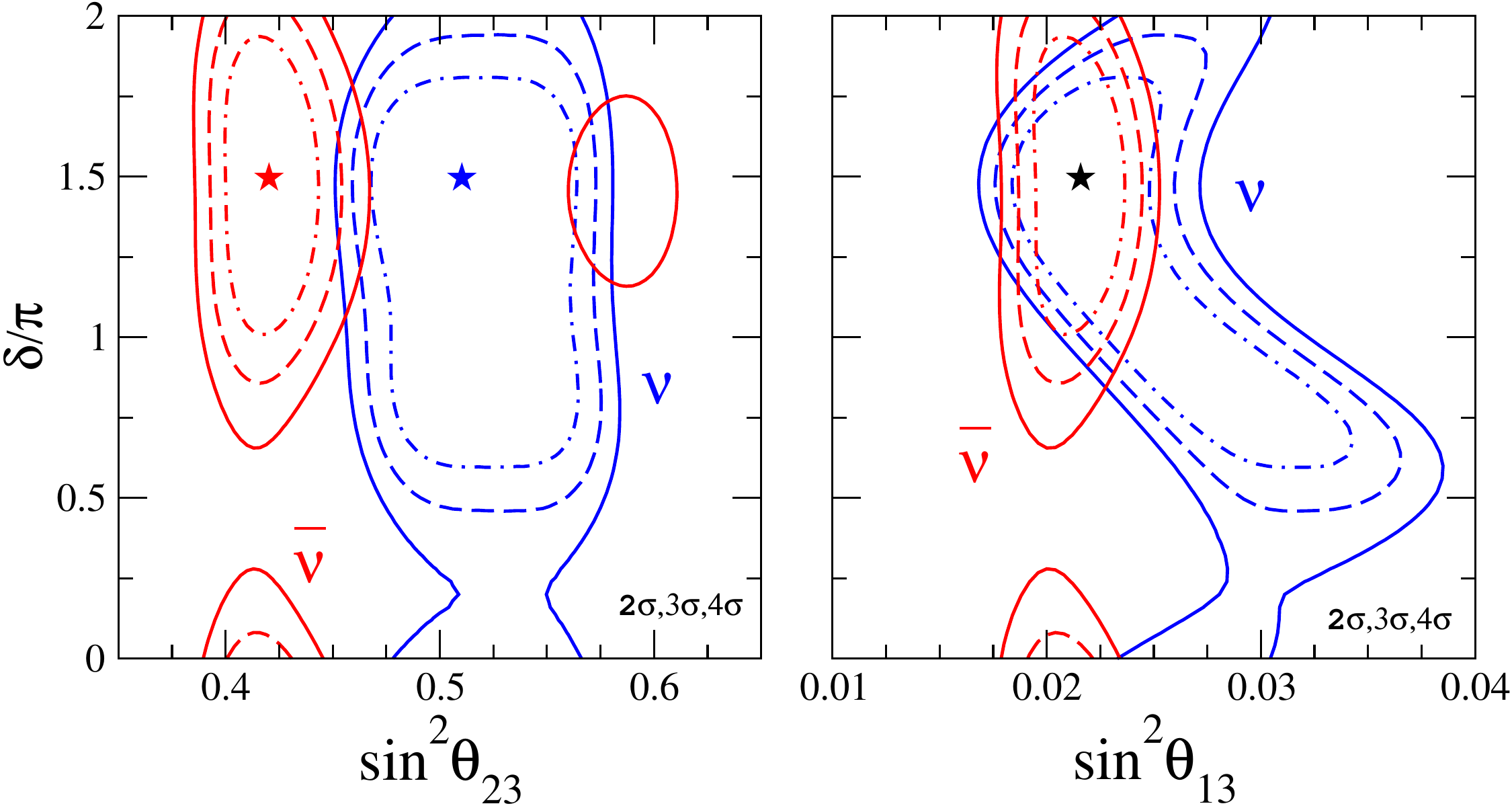}
        \captionsetup{justification=raggedright}
        \caption{DUNE expected regions at 2$\sigma$, 3$\sigma$ and 4$\sigma$ in the $\sin^2\theta_{23}$ -- $\delta$ plane (left) and 
        the $\sin^2\theta_{13}$ -- $\delta$  plane (right) for neutrinos (blue) and antineutrinos (red). The stars indicate the assumed true values for neutrino (blue) and antineutrino (red) oscillation parameters and for both (black) in the right panel.}
	\label{fig:sq23-sq13-del}
\end{figure}

\begin{figure}[h!]
 \centering
        \includegraphics[width=0.9\textwidth]{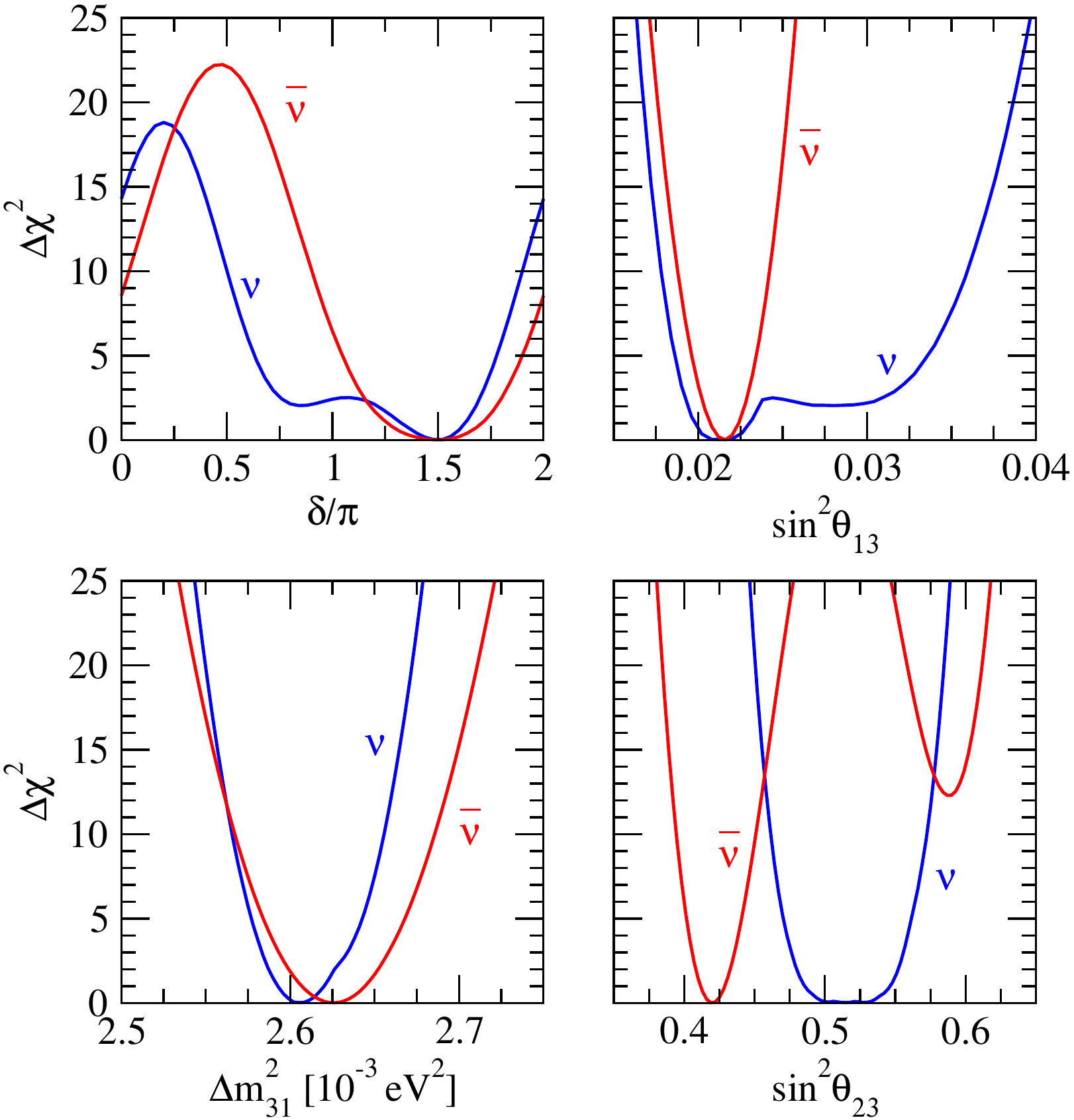}
       \captionsetup{justification=raggedright}
        \caption{$\Delta\chi^2$ profiles as a function of $\delta$, $\sin^2\theta_{13}$, $\Delta m^2_{31}$ and $\sin^2\theta_{23}$ 
        from the separate analysis of neutrino and antineutrino data in DUNE assuming the atmospheric parameters as determined by 
        the separate analysis by the T2K collaboration.}
	\label{fig:sensitivity-t2k}
\end{figure}

Note also that the antineutrino run in DUNE alone could resolve 
the octant of the atmospheric mixing angle at 3$\sigma$  if $\sin^2\overline{\theta}_{23}=0.42$ turns out to be the true value. 
 In the right panel of Fig.~\ref{fig:sq23-dm31-sq13} we see that the DUNE  
neutrino mode is not very sensitive to the reactor angle $\theta_{13}$, since values as large as $\sin^2\theta_{13}=0.034$ 
(far from the Daya Bay upper bound) are allowed
at the 2$\sigma$ level.
On the other hand,  in the right panel of Fig.~\ref{fig:sq23-sq13-del} one can see the impact of considering a prior on $\sin^2\overline{\theta}_{13}$  
on the determination of $\overline{\delta}$.
The lack of a prior in the neutrino mode results in a more reduced sensitivity to the CP--phase $\delta$. 
This can also be observed in Fig.~\ref{fig:sensitivity-t2k}, where we plot the $\Delta\chi^2$ profiles of the oscillation parameters.
On the contrary, the neutrino mode is more sensitive to the atmospheric mixing angle and  mass splitting, see lower 
panels in Fig.~\ref{fig:sensitivity-t2k}.

As commented above, the main constraint on the reactor angle in antineutrino oscillations comes from the  Daya Bay experiment~\cite{An:2016ses}, 
while in the case of neutrino oscillations no such measurement exists. 
So, even though the neutrino channel has higher statistics because of the reduced cross section for antineutrinos, the constraints from 
Daya Bay improve drastically the sensitivity in the antineutrino channel.
We will see in  Sec.~\ref{sec:sensitivity} that a good determination of the reactor angle also helps 
in resolving the octant problem of the atmospheric angles. 
For example, when we study  
the sensitivity of DUNE to the atmospheric angle for two different true values of it, we will see
that in the neutrino mode the degenerate solutions cannot be distinguished 
from the real solution, but in antineutrino mode it is disfavored at more than 3$\sigma$ confidence level, due to the good determination 
of $\overline{\theta}_{13}$.

\section{DUNE sensitivity to CPT--violating neutrino oscillation parameters}
\label{sec:sensitivity}

\begin{table}[ht!]\centering
  \catcode`?=\active \def?{\hphantom{0}}
   \begin{tabular}{lc}
    \hline
    parameter & value 
    \\
    \hline
    $\Delta m^2_{21}$& $7.56\times 10^{-5}\eVq$\\  
    $\Delta m^2_{31}$&  $2.55\times 10^{-3}\eVq$\\
    $\sin^2\theta_{12}$ & 0.321\\ 
     $\sin^2\theta_{23}$ &  0.43, 0.50, 0.60\\
    $\sin^2\theta_{13}$ & 0.02155\\
   $\delta$ & 1.50$\pi$\\
       \hline
     \end{tabular}
       \captionsetup{justification=centering}
     \caption{Oscillation parameters used to simulate neutrino and antineutrino data analyzed in Sec.~\ref{sec:sensitivity}.}
     \label{tab:par2} 
\end{table}

\begin{figure}[ht!]
 \centering
        \includegraphics[width=0.9\textwidth]{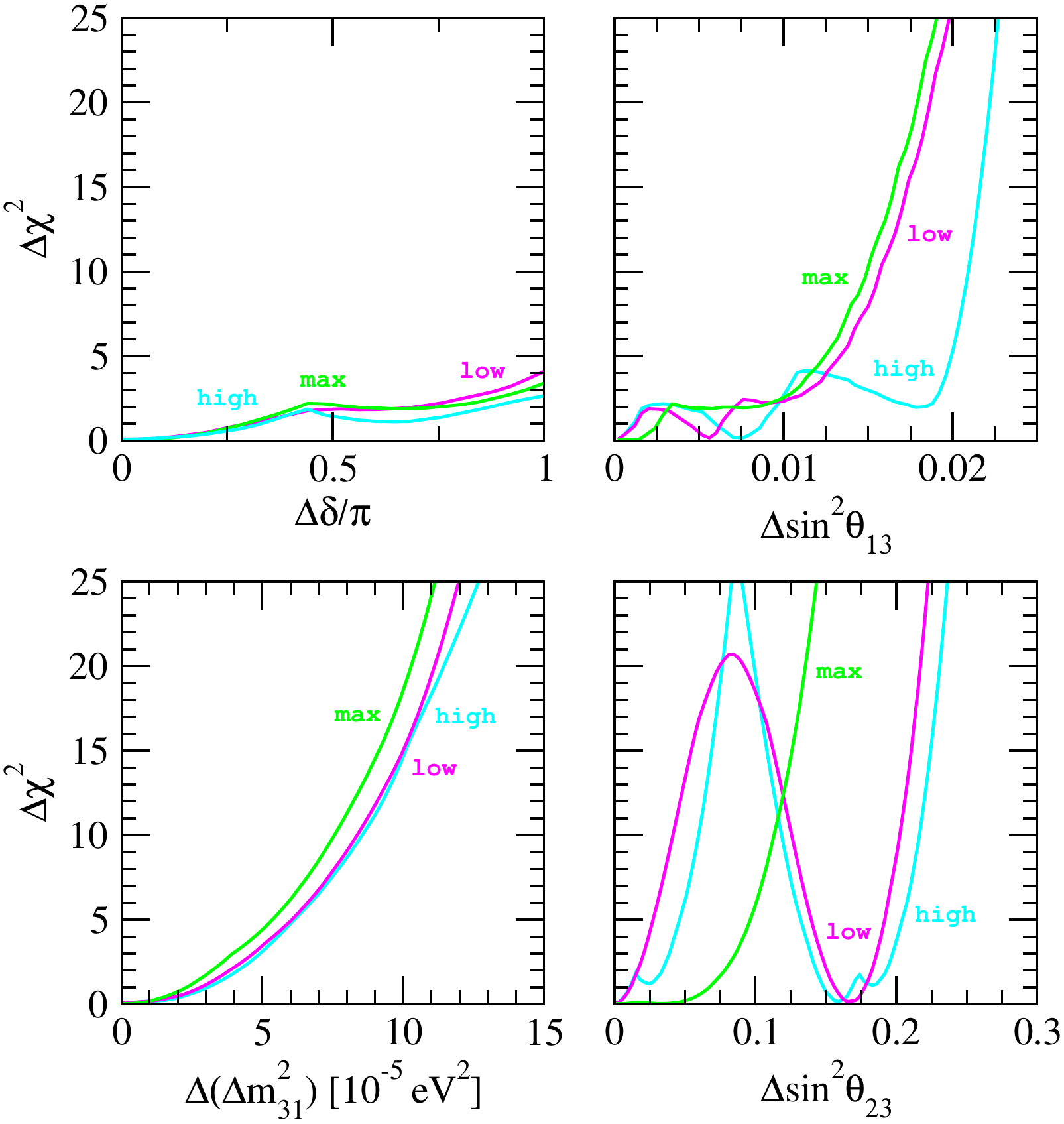}
        \caption{The sensitivities of DUNE to the difference of neutrino and antineutrino parameters: 
        $\Delta\delta$, $\Delta(\Delta m_{31}^2)$, $\Delta(\sin^2\theta_{13})$ and $\Delta(\sin^2\theta_{23})$  
        for the atmospheric angle in the lower octant (magenta line),  in the upper octant (cyan line) and for maximal mixing (green line).}
	\label{fig:sensitivity-CPT}
\end{figure}

In this section we study the sensitivity of DUNE to measure CPT violation in the neutrino and antineutrino oscillation parameters.
For a given oscillation parameter $x$,  we first perform  simulations of the DUNE experiment with $\Delta x = |x-\overline{x}| = 0$, i.e., 
assuming equal parameters for neutrinos and antineutrinos. Next, we estimate the sensitivity of DUNE to $\Delta x\neq 0$. 
In our analysis of the DUNE neutrino and antineutrino mode, we vary freely all the oscillation parameters except the solar ones, as explained above. 
The treatment of the reactor angle $\overline{\theta}_{13}$ in the case of the antineutrino mode is also slightly different, since we put the same prior 
on $\sin^2\overline{\theta}_{13}$ as in the previous section.
To simulate the  data in DUNE we consider as true parameters the values in Table \ref{tab:par2}. To explore  possible correlations 
between DUNE CPT sensitivity and the atmospheric octant, we have chosen three values for $\theta_{23}$.
 First we choose  its best fit value as given in Ref.~\cite{deSalas:2017kay}, which lies in the lower octant. Then,  we 
also consider $\theta_{23}$ in the upper octant as well as maximal atmospheric mixing.
 After minimizing over all parameters except $x$ and $\overline{x}$, we calculate 
\begin{equation}
 \chi^2(\Delta x) = \chi^2(|x-\overline{x}|) = \chi^2(x)+\chi^2(\overline{x}),
 \label{eq:chi2-nu-nubar}
\end{equation}
where we have considered all possible combinations of $|x-\overline{x}|$. Our results are presented in Fig.~\ref{fig:sensitivity-CPT}, 
  where we plot three different lines, labelled as "high", "max" and "low". These refer to the assumed value for the atmospheric angle:
 in the lower octant (low),  maximal mixing (max) or in the upper octant (high).
There, one can see that there is no sensitivity to $\Delta(\sin^2\theta_{13}) = |\sin^2\theta_{13} - \sin^2\overline{\theta}_{13}|$, 
nor to $\Delta\delta = |\delta - \overline{\delta}|$.
 Note that, in the case of $\Delta(\sin^2\theta_{13})$, there would be a $3\sigma$ exclusion only for 
 $\Delta(\sin^2\theta_{13})\approx 0.015$, which is basically 
of the order of $\sin^2\overline{\theta}_{13}=0.02155$. For $\Delta\delta$ we would not even disfavor any value at more than $2\sigma$ 
confidence level. 

\begin{figure}[ht!]
 \centering
        \includegraphics[width=0.9\textwidth]{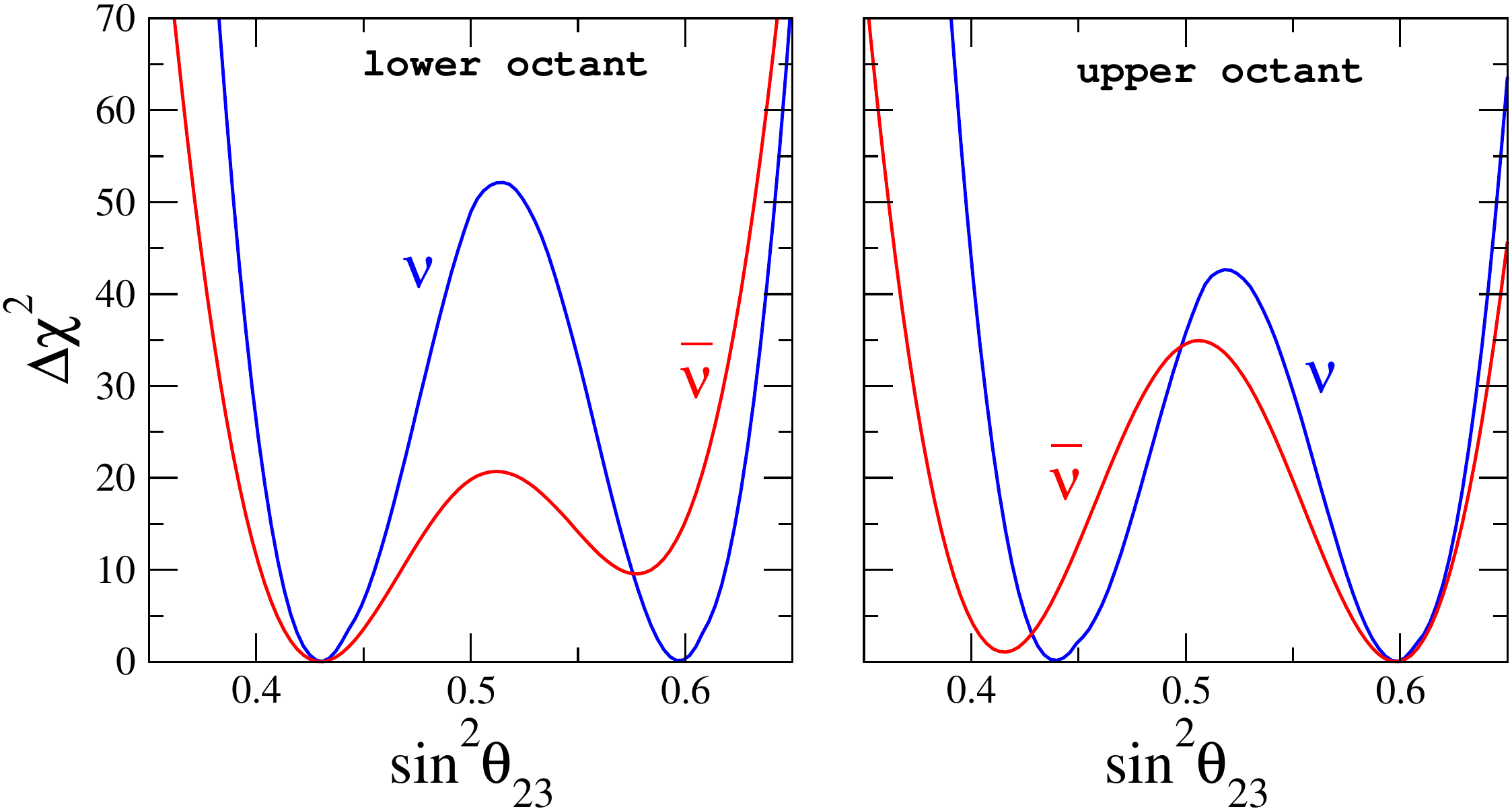}
        \caption{The sensitivity to the atmospheric angle for true values in the lower (left) and upper (right) octant}
	\label{fig:sens-sq23}
\end{figure}

On the contrary, we obtain very interesting results for $\Delta(\Delta m_{31}^2)$ and $\Delta(\sin^2\theta_{23})$. 
First of all, we find that DUNE should be able to set bounds on $\Delta(\Delta m_{31}^2)$ tighter than $8.1\times 10^{-5}$ at 
$3\sigma$ confidence level. 
This would imply an improvement of one order of magnitude  with respect to the old bound in Ref.~\cite{Adamson:2013whj} and four
orders of magnitude with respect to the neutral Kaon bound, once it is viewed as a bound on the mass squared.
Concerning the  atmospheric mixing angle, we obtain different results depending on the  true value assumed to simulate DUNE data. 
In the lower right panel of Fig.~\ref{fig:sensitivity-CPT} we see the different behaviour obtained for maximal  $\theta_{23}$ 
and $\theta_{23}$  in the upper or lower octant.
 In the case of true maximal mixing, the sensitivity  increases with $\Delta(\sin^2\theta_{23})$, as one might expect. 
However, if we assume the true values to be in the first or second octant,  a degenerate solution appears in the 
complementary octant, as can be seen in 
Fig.~\ref{fig:sens-sq23}. 
Since there is no prior on $\sin^2\theta_{13}$ in the neutrino mode, the second fake solution  survives with 
$\Delta\chi^2\approx 0.15$. Hence, in minimizing over 
$|\sin^2\theta_{23} - \sin^2{\overline{\theta}}_{23}|$,  a second minimum appears if one value is in the lower octant and the 
other one in the upper one close to the degenerate solution. This means that, if in nature for example $\sin^2\theta_{23}\approx0.43$ 
and $\sin^2\overline{\theta}_{23}\approx0.60$, DUNE would be blind to this difference,  as long as no better determination of $\theta_{13}$ is 
obtained. This behavior can be explained by looking at the $\Delta\chi^2$ profiles 
of the atmospheric angles in Fig.~\ref{fig:sens-sq23}. 
Note that the neutrino channel alone is basically blind to the octant 
discrimination and then the degenerate solution always appears. Even in the antineutrino channel, the degeneration disappears only if $\sin^2\overline{\theta}_{23}$ 
lies in the lower octant. If it lies in the upper octant, the degenerate solution also shows up. This is because the constraint 
on $\sin^2\overline{\theta}_{13}$ from Daya Bay pulls $\sin^2\overline{\theta}_{23}$ into the lower octant. Hence, both solutions appear also in this case.
Note also that, in the cases considered here, every single channel on its own could rule out maximal mixing at 4$\sigma$ - 7$\sigma$ confidence level. 

\section{Obtaining imposter solutions}
\label{sec:impost}

In neutrino experiments whose beam is produced at accelerators, neutrino and antineutrino data are obtained
on separated runs. However, courtesy of the smallness of antineutrino cross section as compared to the neutrino one, roughly 
only one third of the data are obtained with the former, implying larger statistical errors. Because of that and under the 
seemingly "light" assumption of CPT conservation, it is tempting to perform a joint analysis. Such a path, as we have shown so far, is 
not risk-free. First of all, the opportunity to set the best limit on the possible CPT violation in the mass-squared of
elementary particles and antiparticles is lost. And most important, if CPT is violated in Nature, the gain in
statistics is done by sacrificing the physics. The outcome of the joint data analysis will not be that of either 
channel but what we call an imposter solution. A solution which results from the combined analysis but does
not correspond to the true solution of either channel.

Nevertheless, in experiments and also global fits one normally assumes CPT to be conserved. 
In this case the $\chi^2$--functions are computed according to
\begin{equation}
 \chi^2_{\text{total}}=\chi^2(\nu)+\chi^2(\overline{\nu})
\end{equation}
before marginalizing over any of the parameters. In contrast, in Eq.~\ref{eq:chi2-nu-nubar} we first marginalized 
over the parameters in neutrino and antineutrino mode separately and then added the marginalized profiles. 

\begin{figure}[ht!]
 \centering
        \includegraphics[width=0.6\textwidth]{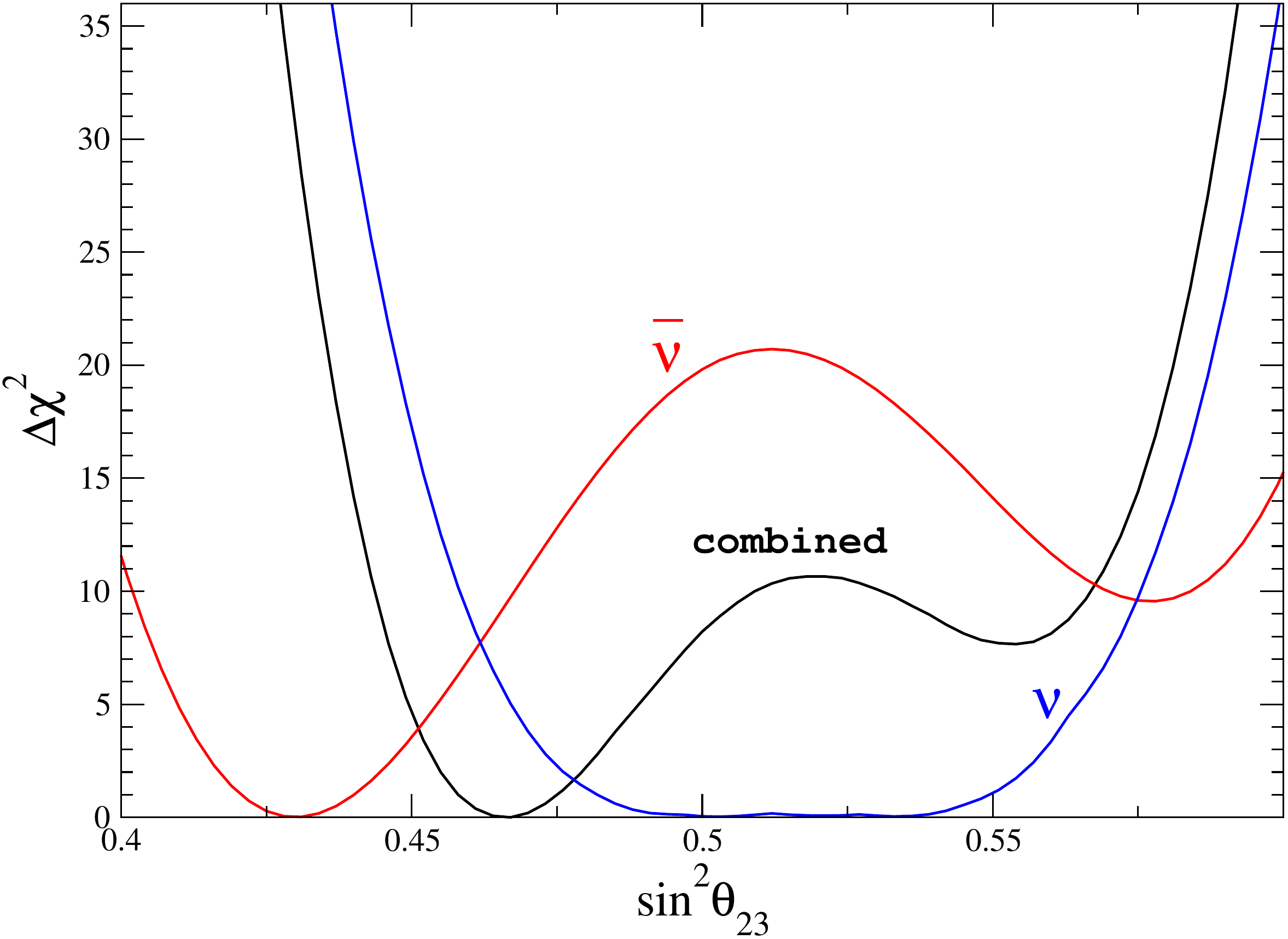}
        \caption{DUNE sensitivity to the atmospheric angle for neutrinos (blue), antineutrinos (red) and to the combination of both under the assumption of CPT conservation (black).
         }
	\label{fig:imposter-sq23}
\end{figure}

In this section we assume that CPT is violated, but treat our results as if it was conserved. 
We assume that the true value for atmospheric neutrino mixing is $\sin^2\theta_{23}=0.5$, while the antineutrino mixing angle is given by  
$\sin^2\overline{\theta}_{23}=0.43$. The remaining oscillation parameters are fixed to the values in Tab.~\ref{tab:par2}. 
If we now combine the results of our simulations for these values, but assume the same mixing for neutrinos and antineutrinos in the reconstruction analysis, we obtain
the sensitivity to the atmospheric angle presented in Fig.~\ref{fig:imposter-sq23}. We also plot the individual reconstructed profiles  for neutrinos and antineutrinos for comparison.
By combining the two results we obtain the best-fit value at $\sin^2\theta^\text{comb}_{23}=0.467$, disfavoring the true values at close
to 3$\sigma$ and more than 5$\sigma$ for neutrino and antineutrino parameters, respectively. 

\begin{figure}[ht!]
 \centering
        \includegraphics[width=0.9\textwidth]{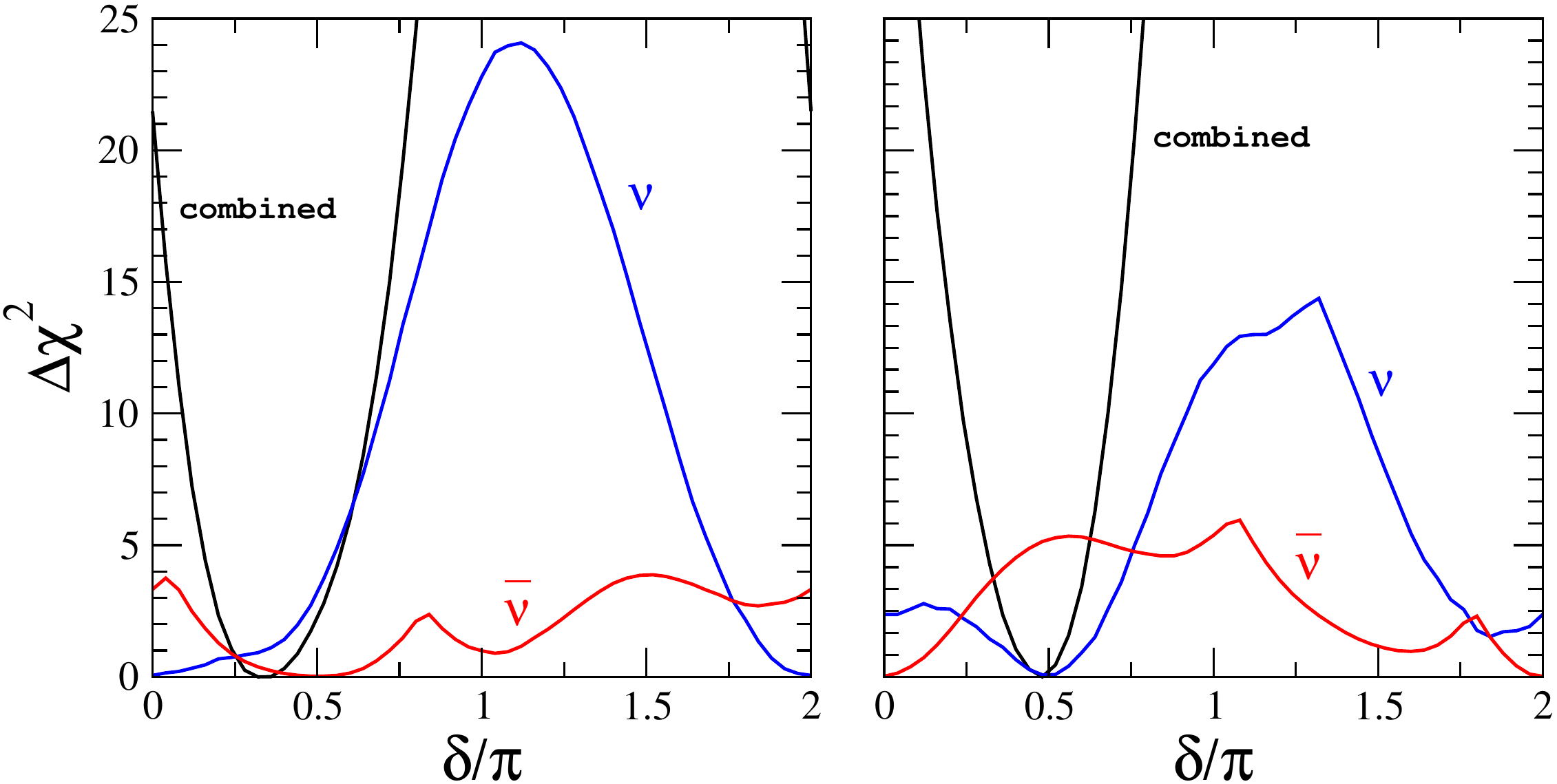}
          \caption{Obtaining imposter solutions for the CP-phases. As before, blue (red) line corresponds to DUNE sensitivity for neutrinos (antineutrinos). Black line corresponds to the combined neutrino + antineutrino sensitivity if CPT conservation is assumed. }
	\label{fig:imposter-delta}
\end{figure}

We also performed a similar study fixing $\sin^2\theta_{23}=\sin^2\overline{\theta}_{23}=0.430$, but choosing $\delta = 0 (0.5\pi)$ and 
$\overline{\delta}=0.5\pi ( 0)$. The results are presented in the left (right) panel of Fig.~\ref{fig:imposter-delta}. As it can be seen, the good 
sensitivity to $\overline{\delta}$ we found before holds only if $\overline{\delta}\approx1.5\pi$, due to correlations with $\overline{\theta}_{13}$. 
Putting $\overline{\delta} = 0$ or $0.5\pi$ all sensitivity gets lost. On combining both channels the value of $\delta=0$ gets highly disfavored,
close to 5$\sigma$ in one case and more than 7$\sigma$ in the other, even though it is one of the true values of oscillations.

\section{Summary and conclusions}
\label{sec:summary}
Being the CPT theorem one of the few solid predictions of local relativistic quantum field theories, the implications
of its potential violation cannot be underestimated. If found, CPT violation will threaten the very foundations
of our understanding of Nature. The impressive limit on CPT from the neutral Kaon system, normally referred as the
world best bound on CPT, turns out to be very weak if viewed as constraint on the mass squared. For a true test one
has to turn to neutrino oscillation experiments where more than four orders of magnitude better limits can be obtained. 
 DUNE therefore has the capability of obtaining the best limit on the possible CPT violation in mass-squared of
particles and antiparticles testing the region where a potential CPT violation arising from non-local quantum gravity, 
which is suppressed by Planck scale, is well within reach. 

It should be noticed that, due to the current tension between  KamLAND results (using antineutrinos) and 
solar neutrino experiments (using neutrinos), bounds on the solar mass difference, {\it i.e.}, $\Delta(\Delta m^2_{21})$
are not better than the ones obtained with future DUNE data for the atmospheric splitting. 
It is also worth noticing that  since the Daya Bay prior is only applicable on $\overline{\theta}_{13}$, contrary to the general case, the CP violating phase 
sensitivity improves in this mode as compared to the neutrino one.

In summary, regardless of whether the atmospheric mixing angle is in the lower octant, the upper one or it is just maximal
mixing, DUNE will test CPT violation in the atmospheric mass difference to an unprecedented level,  being able to
place a bound (if not finding it)
\begin{equation}
|\Delta m_{31}^2 - \Delta\overline{m}_{31}^2| < 8.1 \times 10^{-5} \;\; \mbox{eV}^2
\end{equation}
at 3$\sigma$ C.L. Four orders of magnitude more stringent than the neutral Kaon mass  difference, once written in this form.

As we have explicitly shown, the separate analysis of the neutrino and antineutrino runs is not an option. Imposter 
solutions crop up in the joint analysis which do not capture the physics of either  mode.

\section{Acknowledgments}

GB acknowledges support from the MEC and FEDER (EC) Grants SEV-2014-0398, FIS2015-72245-EXP, and FPA2014-54459 
and the Generalitat Valenciana under grant PROMETEOII/2017/033. GB acknowledges partial support from the European 
Union FP7 ITN INVISIBLES MSCA PITN-GA-2011-289442 and InvisiblesPlus (RISE) H2020-MSCA-RISE-2015-690575. 
CAT and MT are  supported by the Spanish grants FPA2014-58183-P,  FPA2017-85216-P and
SEV-2014-0398 (MINECO) and PROMETEOII/2014/084 and GV2016-142 grants
from Generalitat Valenciana. MT is also supported a Ram\'{o}n y Cajal contract (MINECO).

\begingroup
\raggedright
\sloppy

\bibliographystyle{apsrev}

\end{document}